# A Neuron Based Switch: Application to Low Power Mixed Signal Circuits

Alex Pappachen James, Fayaz Shariff and Akshay Kumar Maan

Queensland Micro- and Nanotechnology Centre,
Griffith University, Nathan Campus Australia,
4111
apj@ieee.org

*Abstract*— **Human brain is functionally and physically complex. This 'complexity' can be seen as a result of biological design process involving extensive use of concepts such as modularity and hierarchy. Over the past decade, deeper insights into the functioning of cortical neurons have led to the development of models that can be implemented in hardware. The implementation of biologically inspired spiking neuron networks in silicon can provide solutions to difficult cognitive tasks. The work reported in this paper is an application of a VLSI cortical neuron model for low power design. The VLSI implementation shown in this paper is based on the spike and burst firing pattern of cortex and follows the *Izhikevich* neuron model. This model is applied to a DC differential amplifier as practical application of power reduction.**

## I. INTRODUCTION

Neurons are the fundamental unit in the human brain that account for intelligence and memory. They vary in their physical structure and function, and are a large part of biological nervous system. Of particular interest are the spiking neurons found in the brain that could hold the key to our understanding of intelligence. Inspired by this idea, over the last decade researchers have come up with various plausible mathematical models to re-engineer the neuronal function [1]–[23].

The 'integrate and fire model' of a neuron was one of the first methods to be implemented in hardware. The circuit model consists of a combination of capacitors and resistors [24]. A direct implementation of a capacitor and a resistor in an integrated circuit can lead to high power consumption and can consume large areas on the chip. Instead, we can use Metal-Oxide-Semiconductor Field Effect Transistor (MOSFET) to simulate a capacitor and a resistor via the ideas of equivalent resistor and capacitor models of MOSFET. Following this principle, in the past, optimized VLSI cortical neuron structures were designed by employing *Izhikevich* computational model [2] and 'integrate and fire model'. Similar works that have been reported in the past that follow this approach are: Hindmarsh-rose neuron model [24], Volterra system [25], Fitzhugh-Nagumo neuron model [26], Morris- Lecar neuron model [27] and Oregonstor model [28]. Most of these model circuits consume over 20 transistors and focus largely on the issues of low power. However, as technology parameters change, these circuits lose the accuracy of modelling the time signals compared to a biological neuron. Hardware implementation of cortical neurons with spiking and bursting behaves very similar to biological neurons. This paper is an extension to the work done by [1], wherein MOSFET based implementation using Izhikevich computational model [2] is shown and analysed.

The spikes (pulses) generated by the neuron circuit are used in an application of a low power DC-differential amplifier. Today, power consumption is considered as one of the most important problems in high performance in MOSFET circuits. As the component density increases, the need for lower power consumption increases [29]. Traditionally two methods have been employed to minimise power consumption in MOSFET amplifiers: (1) reducing supply voltage, and (2) novel ways to control the bias current. However, since the MOSFET noise is dependent on current, reduction in current beyond a certain limit is not recommended for high performance analog circuits. On the other hand, the drawback of reducing power supply in the sub-micron MOSFET technologies is that it results in a reduced signal swing and low dynamic range. This paper presents a novel method of reducing power dissipation of the differential amplifier by using mixed signal approach to rapidly switch ON and OFF the internal circuitry. When the duration in the switched OFF condition increases as compared to the duration when the circuit is switched ON, the overall power dissipation of the circuit decreases. The DC output thus obtained is a sampled DC output signal in time.

## II. BACKGROUND

The cortical neuron circuit reported by [1] consists of 14 MOSFET transistors as shown in Fig. 1. The circuit consists of two state variables: (1) a membrane potential circuit and (2) a slow variable circuit. The membrane potential and slow variable circuits are biased using current mirrors. M11 and M12 show the implementation of differential amplifier and M10 and M9 show the implementation of current mirror circuit. Transistors M9 and M10 produce a high gain and are used to balance the load resistances. This circuit has low noise, high speed and low power consumption; these characteristics are largely attributed to the effective use of pull up and pull down MOSFETs that are connected along the power supply rails. The output of the circuit is dependent upon the ratios of $C_u$ and $C_v$ (Fig. 1), and equivalent capacitance and resistance of the MOSFETs (to model part of the 'integrate and fire') connected to the post synaptic input node. Experimental test results are obtained with SPICE 0.35μm MOS technology. A post synaptic current of 0.1μA is applied at the input. The circuit can be verified using various firing patterns such as: low threshold spiking, chattering, intrinsic bursting, fast spiking and regular. Different spiking and bursting patterns depend up on the variable voltages Vc and Vd. It is also possible to obtain different firing pattern by changing the length and width of transistors M4, M6 and M7. Various levels of spiking and bursting can occur by changing the ratios of Cu and Cv, and the length and width of transistors M6 and M7. It may be noted that VLSI cortical neuron firing patterns have a higher frequency of firing in the range of micro seconds in comparison with millisecond range of biological neuron. In our investigation we focus on fast spiking and bursting behaviours of neuron model. In the brain, neurons communicate with each other using short electrical pulses called spiking or action potential. There are several types of spiking and bursting patterns in the brain. All excitatory cells are divided in to four classes [2]: RS (regular spiking), IB (intrinsically bursting), and CH (chattering). All inhibitory cortical cells are divided in to two classes [2]: FS (fast spiking) and LTS (low threshold spiking). There are also some other types to provide input to the cortex: TC (thalamo-cortical) and RZ (resonator). Some examples of firing patterns we use in this paper are shown in Fig. 2.

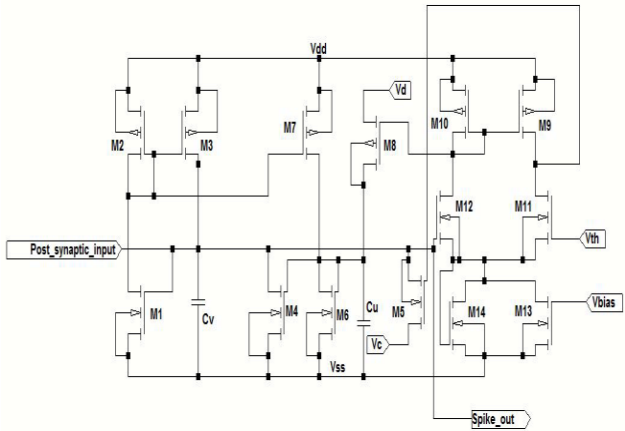

Fig. 1. The cortical VLSI neuron circuit reported by [1]. The MOSFETs in the circuit is used to implement the resistor (R) and capacitor (C) required for implementing the Izhikevich model [1].

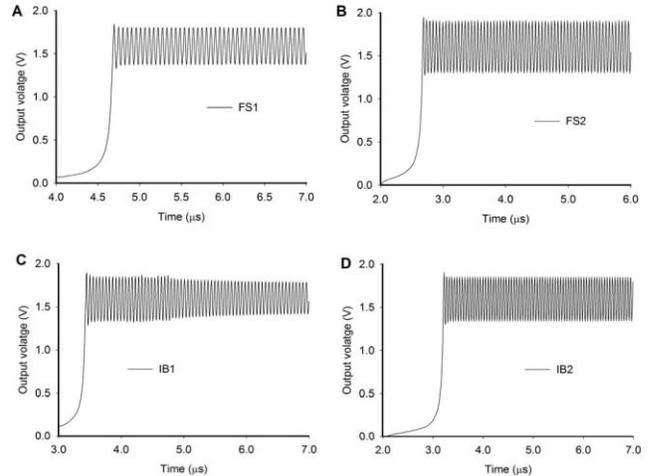

Fig. 2. The graphical illustration of the cortical neuron firing patterns for fast piking (FS) and intrinsically bursting (IB) neuron specifications simulated using the VLSI cortical neuron model described in Fig.1. (A) and (B) shows the firing pattern for simulated fast spiking neuron, (C) and (D) shows the iring pattern from the simulated intrinsic bursting (IB) neuron.

## III. PROPOSED DESIGN

Fig. 3 shows a modified single stage differential amplifier implementation. In this circuit, S1 and S2 are the MOSFET switches that are implemented using transmission gate logic. M3 and M4 form the differential amplifier pair. V1 and V2 are the test inputs, and $V_{out}$ is the single ended output. M1, M2, M5, and M6 form the current mirror bias to the differential amplifier. The switches were driven by pulses generated by the neural circuit. A high pulse turned the switch ON and the circuit was charged up, while a low pulse turned the switch OFF which in turn cut-off the power supplies to the internal circuits. By adjusting the duty cycle of the pulses and using the various modes of operation of the cortical neuron circuit, we attempted to save power dissipation of the circuit during OFF state of the switches. As the high pulse became narrower and low pulse became wider, the amplification was ON for a much smaller time than it was OFF. To compensate for the distortions that could occur through such discrete amplification, the frequency of the pulse was kept very high, which is naturally obtained from cortical neuron models.

A MOSFET circuit dissipates power via two means: dynamic and static. Dynamic power dissipation occurs due to charging and discharging of the equivalent MOSFET capacitances in the circuit. When the MOSFET is ON, in differential amplifier, the approximate value of effective capacitance can be shown as:

$$C = \frac{2 \in A}{3t_{ox}} \qquad (1)$$

Hence the maximum possible capacitance between the two supplies would be the number of MOSFETs times the effective capacitance of each MOSFET. Since under normal conditions there are a total 6 MOSFETs in the circuit as shown in Fig. 3, we have:

$$P_s = 6I_{BIAS}(V_{dd} + V_{ss}) \qquad (5)$$

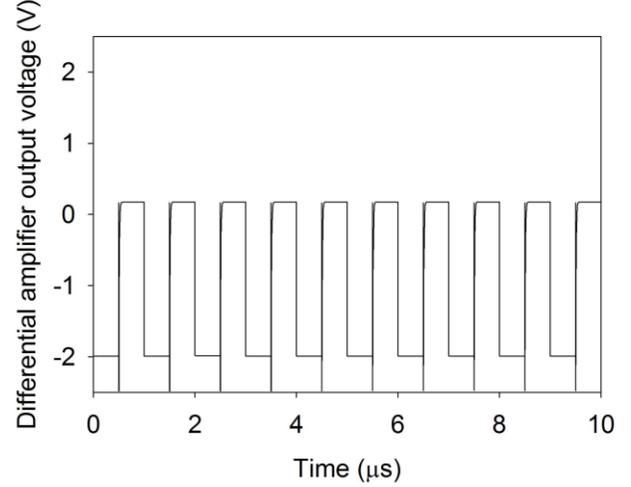

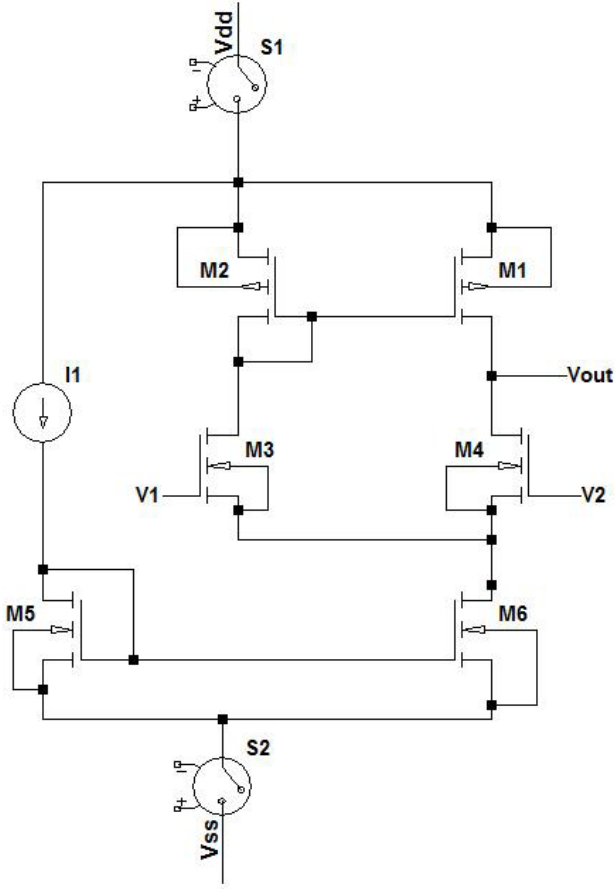

Fig. 3. The proposed circuit schematic illustration of the low-power design. S1 and S2 are the MOSFET switches. M3 and M4 form the differential amplifier pair. VI and V2 the test inputs, and Vout the single ended output. M1, M2, M5, and M6 form the current mirror bias to the differential amplifier.

$$C_{eff} = \frac{4 \in A}{t_{ox}} \qquad (2)$$

The dynamic power dissipation is given by:

$$P_d = f \cdot \frac{\in A}{t_{ox}}(V_{dd} + V_{ss})^2 \qquad (3)$$

where $f$ represents frequency of operation.

The circuit dissipates static power through leakage current that is a characteristic of MOSFETs since they are not ideal devices. Maximum static power consumed by any MOSFET can be given as:

$$P_{1s} = I_{BIAS}(V_{dd} + V_{ss}) \qquad (4)$$

where, $I_{BIAS\psi}$ is the maximum current (saturation current) flowing through the MOSFETs. Hence, the maximum static power that a circuit can consume is the number of the MOSFETs employed in the circuit times the maximum static power dissipated by any MOSFET. Again, since there are 6 MOSFETs in the circuit of Fig. 3, we have:

Fig. 4. The graphical illustration shows the output voltage of the differential amplifier when a spike of $2\mu s$ is used for controlling the switches.

From the discussion above, we can write the maximum total power dissipation in a differential amplifier under normal circumstances with 6 MOSFETS to be:

$$P_{n,total} = 6[f \cdot \frac{2 \in A}{3t_{ox}}(V_{dd} + V_{ss})^2 + I_{BIAS}(V_{dd} + V_{ss})] \qquad (6)$$

When we introduced a simple switch between the power supply and the internal circuit as shown in Fig. 3, the total power dissipation for the circuit when the switches were closed would be:

$$P_{s,total,ON} = (6+n)[f \cdot \frac{2 \in A}{3t_{ox}}(V_{dd} + V_{ss})^2 + I_{BIAS}(V_{dd} + V_{ss})] \qquad (7)$$

Here, it can be noted that there are $6 + n\ \psi$MOSFETs in total, including $n$ MOSFETs to account for the design of the switch. However, when the spikes go low, the circuit is OFF and the total theoretical power dissipation of the circuit is 0:

$$P_{s,total,OFF} = 0 \qquad (8)$$

As an example, for the spikes of 10MHz and duty cycle of 1%, the circuit dissipates $P_{s,total,ON}$ for $1ns$ and dissipates no power for $99ns$ in one time cycle. Hence the design allows power dissipation for only 1% of the time the differential amplifier is amplifying a signal, thus saving approx. 99% of the total power dissipated by it under non-switching conditions. Using the cortical neuron, we generated spikes with $2\mu\ s$ period to control the ON-OFF cycles of the switches. The amplifier is given a differential input voltage of $2ms$ and the resulting output is shown in Fig 4. The gain of the amplifier is designed for a value of 1000, making the expected output value to be in the proximity of 2V. Figure 4 shows that the circuit remains OFF when the switch is closed and shows the desired output with proper amplification during the period when the switch is ON.

## IV. DISCUSSION AND CONCLUSION

The paper presented a practical application of the cortical VLSI neuron model in reducing power consumption of redundant circuits. As evident from the analysis of power dissipation in the previous section, an increased value of $n\psi$ increased the power consumption during the ON periods of the switches. This means it could negate the savings if the switch is permanently connected. However, in practice, the performance on power savings will be on a positive note as the number of internal components in a target redundant circuit is more than that of a switch. Another issue was the dynamic power, which is dependent on the frequency and internal capacitances. High frequency switching have an effect on the slew rate of the differential amplifier, but it can be addressed by designing additional stages such as buffer circuits and cascade amplifiers. Such design techniques are required to compensate for the parasitic capacitance in classical devices under sub-nano meter channel lengths. The example of a single stage differential amplifier that uses 6 transistors may be a naive approach, however, it provides a general method applicable to a wide range of practical applications in real-time mixed-signal processing systems. Such a premise can be extended to more complex circuits having an array of transistors, such as memory cells, signal processing circuits and image processing circuits.